\documentstyle[12pt]{article}
\begin{document}
\baselineskip=7.5mm
\begin{titlepage}
\begin{flushright}
  KOBE-TH-95-06\\
\end{flushright}
\vspace{2.3cm}
\centerline{{\large{\bf Quaternionic Mass Matrices}}} 
\centerline{{\large{\bf and}}}
\centerline{{\large{\bf CP Symmetry}}}
\par
\par
\par\bigskip
\par\bigskip
\par\bigskip
\par\bigskip
\par\bigskip
\renewcommand{\thefootnote}{\fnsymbol{footnote}}
\centerline{{\bf C.S. Lim}\footnote[1]{e-mail:lim@hetsun1.phys.kobe-u.ac.jp}}
\par
\par\bigskip
\par\bigskip
\centerline{Department of Physics, Kobe University, Nada, Kobe 657, Japan} 
\par
\par\bigskip
\par\bigskip
\par\bigskip
\par\bigskip
\par\bigskip
\par\bigskip
\par\bigskip
\par\bigskip
\par\bigskip
\centerline{{\bf Abstract}}\par

A viable formulation of gauge theory with extra generations in terms of 
quaternionic fields is presented. For the theory to be acceptable, the 
number of generations should be equal to or greater than 4. The 
quark-lepton mass matrices are generalized into quaternionic matrices. 
It is concluded that explicit CP violation automatically disappears 
in both strong- and weak-interaction sectors.

\par\bigskip
\par\bigskip
\par\bigskip
\par\bigskip
\par\bigskip
\par\bigskip
\par\bigskip

\end{titlepage}
\newpage

\vspace{0.8 cm}
\leftline{\large \bf 1. Introduction}
\vspace{0.8 cm}

We still have no conclusive argument on the origin 
of quark-lepton masses, flavor mixings and CP violation. 
To understand the origin of these observables seems 
to be almost equivalent to understanding the origin 
of quark-lepton mass matrices. 

Even if the mass matrices were assumed to be real the 
phenomenon of flavor mixing 
would be still possible. The presence of the flavor mixing, of course, 
demands $N_g \geq 2$, with  $N_g$ being the number of generations. 
In realistic 
complex mass matrices, the phenomenon of CP violation becomes possible, 
but only when $N_g \geq 3$ as was first discussed 
by Kobayashi-Maskawa \cite{MKobayashi}.

Then it may be a natural question to ask whether there 
exists a physical observable or a theoretical 
framework which necessitates the presence of 
higher generations, i.e. $N_g \geq 4$. A trivial 
example of such an observable, one may encounter, may be 
the deviation 
from unitarity of the $3 \times 3$ mixing matrix, 
i.e. $\sum_{i=1}^{3}|V_{ai}|^2 < 1$ with 
$V_{ai}$ being the elements of the matrix. 
So far there is no experimental hint suggesting 
higher generations, including such deviation.  
We, thus, would like to rely on theoretical 
considerations and search for some theoretical framework, 
in which $N_g = 4$ is a critical number. 
Apparently, as far as we work in the framework of 
complex mass matrices the critical value of $N_g$ 
is $3$ and any essential change from the Kobayashi-Maskawa 
scheme will not be expected. 

We, therefore, try to generalize 
the complex structure, i.e., we try to formulate the 
theory in terms of quaternionic spinors, whcih are 
obtained by combining ordinary complex spinors in 
pairs; 2 generations correspond to 1 quaternionic spinor for each 
type of quarks and leptons. The mass matrices for quarks and 
leptons can be naturally quaternionic, without contradicting with 
Lorentz and gauge invariances. 
      The main purpose of the present paper is to investigate 
the consequences of such quaternionic mass matrices. More concretely 
we will investigate the following issues, 
which are closely related with the quaternionic property; 
(i) At which $N_g$ does weak CP violation start to occur ?  
(ii) Is a new insight into the strong CP problem obtained ? 
(iii) Are there any restrictions on quark(lepton) 
masses and/or flavor mixings as the result of the 
quaternionic property ? 

We will see that, in spite of the generalization from 
the complex to the quaternionic matrices, explicit CP 
violation actually disappears in the both of strong 
(assuming $\theta_{QCD} 
= 0$) and weak interaction sectors. Thus the 
quaternionic approach necessarily leads to the scenario 
of spontaneous CP violation. We, however, would like to 
emphasize the fact that in our approach the 
``real" mass matrices are not put by hand, but the absence of 
explicit CP violation is automatically guaranteed as the consequence of the
quaternionic property, i.e. guaranteed for arbitrary quaternionic mass
matrices to start with.  
      For such quaternionic approach to 
make sense $N_g$ must be even. Since we already know 
$N_g \geq 3$, this inevitably means $N_g \geq 4$. 
Correspondingly, the mass matrices should be regarded as 
general $(N_g/2)\times(N_g/2)$ quaternionic matrices.

\vspace{0.8 cm}
\leftline{\large \bf 2. A formulation via 6-dimensional theory}
\vspace{0.8 cm}

Let ${\bf H} = a_1 + a_2{\bf i} + a_3{\bf j} + a_4{\bf k}$ be 
an arbitrary quaternion, with imaginary units ${\bf i},{\bf j}$ and 
${\bf k}$ ($a_1$ to $a_4$:real). ${\bf H}$ can be uniquely decomposed 
into 2 complex numbers, according to ${\bf H} = (a_1 + a_2{\bf i}) 
+ (a_3 + a_4{\bf i}){\bf j}$. In this way an arbitrary quaternionic spinor
reduces to 2 complex spinors. 

To formulate a consistent theory with quaternionic fermions, 
so that it reduces to a viable theory with ordinary complex spinors 
in pairs, is a non-trivial task.  In particular, we immediately 
encounter the problem of whether the imaginary unit `` $i$ " appearing 
in the momentum operator, $ p_{\mu}=-i \partial_{\mu}$ in ordinary 
theories, should be regarded as to commute with quaternions or not. 
If that `` $i$ " is identified with ${\bf i}$ in the quaternionic 
imaginary units, it will not commute with quaternions, especially 
not with ${\bf j}$, leading to $[ p_{\mu},{\bf j}] \neq 0$ . 
It, however, means that the momentum operator 
is gereration dependent (multiplying ${\bf j}$ is equivalent to a 
unitary rotation among different generations), which is not 
acceptable for us. 

A safe way to construct a viable theory, avoiding such problem, is to start
from a 6-dimensional Yang-Mills theory which, after (naive) dimensional
reduction, naturally reduces  to a consistent 4-dimensional theory with
even number of generations. The reason why D=6 has some relevance is that
the 6-dimensional 
Lorentz group $SO(1,5)$ is equivalent to $ SL(2,{\bf H})$ 
where {\bf H} stands for the quaternion \cite{TKugo}, just as 
$SO(1,3)$ is equivalent to $ SL(2,{\bf C})$. A D=6 Weyl fermion can be 
represented as a 2-component quaternionic spinor. $ SL(2,{\bf H})$ 
is also equivalent to $SU^{*}(4)$. A D=6 Weyl fermion, therefore,  can be
equivalently represented as a 4-component complex spinor, and decomposes
into a pair of D=4 Weyl fermions. Apparently, after 
naive dimentional reduction, for instance, where all massive modes are 
neglected, the theory automatically reduces to a viable renormalizable 
D=4 theory. Though utilizing the D=6 theory is quite helpful, once we 
know the way to construct a viable theory we actually can formulate a 
theory with quaternionic fermions in D=4, from the beginning. 
 
The linkage between the theory with quaternionic spinors and the 
theory with ordinary complex spinors can be made through the following
one-to-one correspondence between quaternionic units and $2 \times 2$
complex matrices, which relates the representations in $SL(2,{\bf H})$ and
those in $SU^{*}(4)$ ;

\begin{equation}
1 \leftrightarrow (\begin{array}{cc} 1 & 0 \\ 0 & 1 \end{array}), \   
{\bf i} \leftrightarrow i\sigma_3 = (\begin{array}{cc} i & 0 \\ 
0 & -i \end{array}), \ 
{\bf j} \leftrightarrow i\sigma_2 = (\begin{array}{cc} 0 & 1 \\ 
-1 & 0 \end{array}), \ 
{\bf k} \leftrightarrow i\sigma_1 = (\begin{array}{cc} 0 & i \\ 
i & 0 \end{array}). 
\end{equation}
The assignment above is a bit different from that of Ref.\cite{TKugo}. 
Now regarding the imaginary unit ``$i$" appearing in the momentum 
operator as just ordinary $i$, not as a matrix, the commutativity   
between $i$ and ${\bf j}$ is trivial: 
\begin{equation} 
[i,{\bf j}]= [i, (\begin{array}{cc} 0 & 1 \\ -1 & 0 \end{array})] = 0.
\end{equation} 

One note is in order on the meaning of the ``special" in $SL(2,{\bf H})$.
Usually ``special" means that the determinant of the transformation matrices 
should be 1. In the present case, however, this condition overconstrains 
the matrices, since this condition leaves only $2\times 2\times 4 - 4 
= 12$ real dgree of freedom, while the dimension of $SO(1,5)$ is 
$15$. Thus we have to modify the meaning of the determinant as follows 
for an arbitrary quaternionic matrix ${\bf M}$ \cite{TKugo}: 
\begin{equation} 
det{\bf M} \equiv exp(Tr \ ln{\bf M}), 
\end{equation} 
where the operator $Tr$ should be understood for an arbitrary 
${\bf M}^{'}$ as 
\begin{equation}
Tr{\bf M}^{'} \equiv Re(tr{\bf M}^{'}), 
\end{equation}
where $tr$ means ordinary trace, while $Re$ implies to take only the 
real part.  
Let us note that, in the matrix representation of quaternion 
by the use of Eq.(1), such defined $Tr$ just corresponds to ordinary trace
for the matrix. 
This specific trace, taking the real part, plays an important role 
when we consider CP symmetry of the theory, leading to the 
absence of explicit CP violation. Now that the condition of $det = 1$ reduces 
only one real degree of freedom, as expected. 

The 6-dimensional gamma matrices are given as \cite{TKugo} 
\begin{equation}
\Gamma^M = ( \begin{array}{cc} 0 & \gamma^M \\ \tilde{\gamma}^M & 0 
\end{array}),  \ (M = 0 - 5), 
\end{equation}
where 
\begin{equation}
\gamma^M = ({\gamma}^0, {\gamma}^i),  \tilde{\gamma}^M = ({\gamma}^0, -
{\gamma}^i), 
\end{equation} 
with the $2 \times 2$ quaternionic matrices being given as 
\begin{eqnarray}
{\gamma}^0=(\begin{array}{cc} 1 & 0 \\ 0 & 1 \end{array}), \ 
{\gamma}^1=(\begin{array}{cc} 0 & 1 \\ 1 & 0 \end{array}), \ 
{\gamma}^2=(\begin{array}{cc} 0 & -{\bf i} \\ {\bf i} & 0 \end{array}), 
\nonumber \\ 
{\gamma}^3=(\begin{array}{cc} 1 & 0 \\ 0 & -1 \end{array}), \ 
{\gamma}^4=(\begin{array}{cc} 0 & -{\bf j} \\ {\bf j} & 0 \end{array}), \  
{\gamma}^5=(\begin{array}{cc} 0 & -{\bf k} \\ {\bf k} & 0 \end{array}). 
\end{eqnarray} 
The representation for the gamma matrices given above is in the chiral 
basis, i.e., 
\begin{equation} 
{\Gamma}^7 \equiv - {\Gamma}^0 {\Gamma}^1 \cdot \cdot \cdot {\Gamma}^5 = 
( \begin{array}{cc} I & 0 \\ 0 & -I \end{array}), 
\end{equation} 
where a 6-dimensional Weyl spinor is represented as a 2-componet 
quaternionic spinor, say, the upper half, reducing to a pair of 2-component
complex spinors after 
dimensional reduction. Combining Weyl spinors of diffrent chiralities,
${\bf {\Psi}}_R$ and ${\bf {\Psi}}_L$,  
we get a 4-component full spinor ${\bf {\Psi}}$, ${\bf {\Psi}} = 
{\bf {\Psi}}_R + {\bf {\Psi}}_L$.

 Let us study the Lorentz invariant free lagrangian for a 6-dimensional 
fermion, 
\begin{equation} 
L = Tr(\overline{{\bf \Psi}}{\Gamma}^M i{\partial}_M {\bf \Psi}) 
+ Tr({\bf m} \overline{{\bf \Psi}}_R {\bf \Psi}_L 
+ {\bf m}^{\ast} \overline{{\bf \Psi}}_L {\bf \Psi}_R),  
\end{equation} 
where ${\bf m}$ represents a quaternionic mass. 
We now perform the naive dimensional reduction of this theory into 
4-dimensional world. First we note that a quaternionic fermion 
${\bf \Psi}$ can be uniquely decomposed into two fermions as, 
\begin{equation}
{\bf \Psi} = {\bf \Psi}_1 + {\bf \Psi}_2 {\bf j}, 
\end{equation} 
where each of ${\bf \Psi}_1$ and ${\bf \Psi}_2$ is made of 1 and ${\bf i}$
alone. Under the naive dimensional reduction only the 4-dimensional part of

${\partial}_M$ and therefore ${\Gamma}^M$, which does not contain ${\bf j}$ 
nor ${\bf k}$, survives. Thus the resultant lagrangian reads as 
\begin{equation} 
L = \overline{\Psi}_1 {\gamma}^{\mu}i {\partial}_{\mu}{\Psi}_1 
  + \overline{\Psi}_2 {\gamma}^{\mu}i {\partial}_{\mu}{\Psi}_2 
 + (\begin{array}{cc} \overline{\Psi}_{1R} & \overline{\Psi}_{2R} 
\end{array})\left( \begin{array}[t]{cc} m_1 & - m_2  \\ m_2^{\ast} &
m_1^{\ast} \end{array}\right) \left( \begin{array}[t]{c} {\Psi}_{1L} \\
{\Psi}_{2L} 
\end{array}\right) 
+ h.c. , 
\end{equation}  
where ${\Psi}_{1,2}$ are ordinary complex spinors, and 
${\gamma}^{\mu}$ is an ordinary 4-dimensional 
gamma matrices; both are obtained from the quternionic counterparts, 
${\bf \Psi}_{1,2}$ and ${\Gamma}^{\mu}$ with the quaternionic 
${\bf i}$ being replaced by ordinary imaginary unit $i$. The $2 \times 2$ mass 
matrix, expressed in terms of two complex numbers $m_1$ and $m_2$, 
is just the (transpose of) matrix representation of 
the quaternion ${\bf m}$. This mass matric, though it leads to a 
flavor mixing, has an unacceptable consequence, i.e. degeneracy 
of the mass eigenvalues, $\left|{\bf m}\right|= 
\sqrt{\left|m_1\right|^2 + \left|m_2\right|^2}$. 
This degeneracy may be understood as the reflection of the global 
$SL(1,{\bf H})$ symmetry, independent of Lorentz transformation 
\cite{TKugo}, 
\begin{equation}
 {\bf {\Psi}} \rightarrow {\bf {\Psi}}^{'}={\bf {\Psi}}{\bf u}, 
\end{equation} 
with a unimodular quaternion ${\bf u}, \left|{\bf u}\right|=1$. 
The 4-dimensional spinors $({\Psi}_1, {\Psi}_2)$ behave as 
a doublet under $SU(2)$, which is isomorphic to the $SL(1,{\bf H})$. 
The kinetic term is trivially invariant under this transformation, while 
the mass ${\bf m}$ can be modified into $\left|{\bf m}\right|$. 
We can show that this degeneracy arises for arbitrary even number of 
generations.   

\vspace{0.8 cm}
\leftline{\large \bf 3. A possible alternative formulation}
\vspace{0.8 cm}

We now have to search for an alternative quaternion-like closed algebra 
(division algebra) in order to describe fermions. The almost unique 
choice is to modify the Lie algebra of $SU(2)$, corresponding to the 
imaginary units of quaternion, into the Lie algebra of $SU(1,1)$. Namely,
now the ``imaginary" units ${\bf j}$ and 
${\bf k}$, corresponding to raising and lowering operators of 
$SU(1,1)$, should be accompanied by ordinary $i$: 
\begin{equation}
1 \leftrightarrow (\begin{array}{cc} 1 & 0 \\ 0 & 1 \end{array}),  
{\bf i} \leftrightarrow i\sigma_3 = (\begin{array}{cc} i & 0 \\ 
0 & -i \end{array}),
{\bf j} \leftrightarrow \sigma_2 = (\begin{array}{cc} 0 & -i \\ 
i & 0 \end{array}), 
{\bf k} \leftrightarrow \sigma_1 = (\begin{array}{cc} 0 & 1 \\ 
1 & 0 \end{array}). 
\end{equation}
Accordingly, the gamma matrices are also modified and the 
Clifford algebra, ${\Gamma}^M{\Gamma}^N + {\Gamma}^N{\Gamma}^M = 
2 g^{MN}I_4$ tells us $g_{MN} = diag(1,-1,-1,-1,1,1)$, namely that 
the extra dimensions of the 6-dimensional space-time are 
time-like. So the ordinary conjugate 
$\overline{{\bf {\Psi}}}$ should be replaced by ${\bf {\Psi}}^{\dag} 
{\Gamma}^0{\Gamma}^4{\Gamma}^5$, to ensure the full 6-dimensional 
Lorentz invariance. This conjugate, however, causes an indefinite 
metric ,i.e., 
$\overline{{\psi}_1}{\gamma}^{\mu}i{\partial}_{\mu}{\psi}_1 
- \overline{{\psi}_2}{\gamma}^{\mu}i{\partial}_{\mu}{\psi}_2 $, after 
the dimensional reduction. 
 
Since only 4-dimensional Lorentz invariance is what we really 
have to demand (the Lorentz invariance of the full space-time is 
spoiled under the compactification, anyway), we will just adopt 
ordinary Pauli conjugate, $\overline{{\bf {\Psi}}} = {\bf
{\Psi}}^{\dag}{\Gamma}^0$. 
Thus we have exactly the same form of the free lagrangian as the 
one for the original quaternions, Eq.(9), though the fermions, 
gamma matrices and ${\bf m}$ shoud be understood to be written in terms 
of the modified ${\bf j}$ and ${\bf k}$. We also just follow the 
procedure of the dimensional reduction done above. The only difference 
in the resultant lagrangian for 4-dimensional fermions is that the mass 
matrix is now of the form, 
\begin{equation}
\left( \begin{array}{cc} m_1 &  m_2 \\ {m_2}^{\ast} & {m_1}^{\ast} 
\end{array}\right). 
\end{equation}
We can easily check that the mass eigenvalues are now non-degenerate, 
$\left|m_1\right|\pm \left|m_2\right| $, and the mass matrix deserves to 
the description of the real world.  

\vspace{0.8 cm}
\leftline{\large \bf 4. CP symmetry and the flavor mixing}
\vspace{0.8 cm}

So far we have discussed the free lagrangian for a quaternionic fermion, 
i.e. for two generations. We can immediately generalize the lagrangian 
so that it can contain arbitrary even number of generations, $N_g$, 
with $N_g/2$ quaternionic spinors, ${\bf {\Psi}}_a 
(a = 1$ to $N_g/2)$: 
\begin{equation} 
L = Tr({\overline{{\bf \Psi}}}_a{\Gamma}^M i{\partial}_M {\bf \Psi}_a) 
+ Tr({\bf m}_{ab} \overline{{\bf \Psi}}_{aR} {\bf \Psi}_{bL} 
+ {\bf m}_{ba}^{\ast} \overline{{\bf \Psi}}_{aL} {\bf \Psi}_{bR}). 
\end{equation} 

We will study what does the $(N_g/2)\times (N_g/2)$ matrix ${\bf m}_{ab}$ 
imply concerning flavor mixing and, in particular, CP symmetry, which is of
our main concern. For this purpose, the quaternionic mass materix 
${\bf m}_{ab}$ should be 
replaced by a $N_g \times N_g $ complex matrix, 
obtained by substituting the 
$2\times 2$ matrix form for each quaternionic element according to the 
rule Eq.(13), after the dimensional reduction.  
 
\noindent (i) Flavor mixing 

Concerning the flavor mixing among various generations and its relation 
with mass eigenvalues, we may have, in principle, some new constraint, 
as the quaternionic property restricts the form of the mass matrices to 
some extent, as is seen in Eq.(14). We, however, have not found any 
physically observable constrint. To see the situation, let us investigate the 
simplest case of two generations. In this case, the mass matrix 
of the form of Eq.(14) can be assigned for both of up-type and 
down-type quarks: 
\begin{equation} 
m_U = \left( \begin{array}{cc} m_{u1} & m_{u2} \\ {m_{u2}}^{\ast} 
& {m_{u1}}^{\ast} \end{array} \right), \ 
m_D = \left( \begin{array}{cc} m_{d1} & m_{d2} \\ {m_{d2}}^{\ast} 
& {m_{d1}}^{\ast} \end{array} \right). 
\end{equation}
The mass eigenvalues are non-degenerate as was seen above, 
i.e., $m_{u,c}=\left|m_{u1}\right|\mp \left|m_{u2}\right|$, 
and $m_{d,s}=\left|m_{d1}\right|\mp \left|m_{d2}\right|$. 
The Cabbibo mixing angle turns out to be given as 
\begin{equation}
\theta_C = \frac{1}{2} arg \left(\frac{m_{u1}m_{u2}^{\ast}}{m_{d1}^{\ast}
m_{d2}}\right). 
\end{equation}    
This type of analysis can be easily generalized to higher generation 
cases.

\noindent (ii) Strong CP problem 

One of the interesting solutions to the strong CP problem is to 
assume that the lagrangian does preserve CP symmetry, i.e. no 
explicit CP violation, and therefore that ${\theta}_{QCD} = 0$. 
Then it becomes a 
non-trivial problem to assure the absence of the contribution from 
the flavor dynamics QFD, ${\theta}_{QFD}$, which is expected from the 
spontaneous CP violation in the QFD sector ($\theta = {\theta}_{QCD} 
+ {\theta}_{QFD}$). Nelson and Barr \cite{ANelson} have proposed 
a contrived mechanism 
to guarantee ${\theta}_{QFD} = 0$. It is interesting to note that our 
quaternionic mass matrices, though they contain various phases, 
automatically guarantee ${\theta}_{QFD} = 0$. 
  The proof is quite simple. Namely, in terms of mass matrices for up-type
and down-type quarks, $m_U$ and $m_D$, 
\begin{equation}
{\theta}_{QFD} = arg(det \ m_U) + arg(det \ m_D),
\end{equation}
and each term identically vanishes: 
\begin{equation}
arg(det \ m_U) = arg[exp(Tr \ ln \ m_U)] = arg[exp(Re \ tr \ ln {\bf M}_U)]
= 0, 
\end{equation}
etc., where ${\bf M}_U$ is original $N_g/2\times N_g/2$ quaternionic 
mass matrix for up-type quarks. The presence of the operation $Re$ 
was essential to get this result. We may explicitly check this 
relation for the two generation case, by 
taking the determinant of Eq.(14).

\noindent (iii) Weak CP 

To extract re-phasing invariant measure of weak CP violation, if there is 
any, it is useful in general to analyze the quantities $Im Tr(P_1(H_D) 
P_2(H_U) \cdot \cdot )$ where $P_i$ denote arbitrary monomials of 
hermitian matrices $H_U$ and $H_D$, defined as $H_U \equiv {m_U}^{\dag}m_U$ 
, $H_D \equiv {m_D}^{\dag}m_D$ \cite{Gronau}. In our case we again get no
explicit 
CP violation, since we do not have 
any imaginary part for the arbitrary 
monomials: 
\begin{equation}
Im Tr(P_1(H_D) P_2(H_U) \cdot \cdot ) = Im (Re[tr(P_1({\bf H}_D) 
P_2({\bf H}_U) \cdot \cdot)]) = 0,   
\end{equation} 
with ${\bf H}_{U,D} \equiv {\bf M}_{U,D}^{\dag} {\bf M}_{U,D}$. Thus our
theory is quite different from the Kobayashi-Maskawa 
theory, and inevitably necessitates the mechanism of spontaneous CP 
violation \cite{TDLee}.

The fact that there is no explicit CP violation turns out to be a 
natural consequence of the quaternionic property of the mass matrices. 
Let us note the following fact (similar to the ``reality" of 
the $SU(2)$ representations); 
\begin{equation}
{\bf k} {\bf H}^{\ast \ast} {\bf k} = {\bf H}, 
\end{equation}
where 
${\bf H}^{\ast \ast}$ corresponds to ordinary complex conjugation ${\bf i}
\rightarrow -{\bf i}$, not 
quaternionic conjugation. From this 
we learn that for an arbitrary quaternionic matrix ${\bf M}$, 
$({\bf V}{\bf M} {\bf V}^{\dag})^{\ast \ast} =   
{\bf V}{\bf M} {\bf V}^{\dag}$, with a unitary transformation 
${\bf V} = \frac{1+i{\bf k}}{\sqrt{2}}I$ ($I$: a unit matrix). This means that 
we can move to a basis by the unitary transformation, where 
all mass matrices are real.

While the fermion mass matrices, which result from Yukawa couplings, 
show characteristic features, as discussed above, other interactions 
can be incorporated into the theory just as in ordinary gauge theories. 
This is essentially because the other interactions are ``generation 
blind". One thing we should care about is the scalar potential, since 
we have to device a potential with spontaneous CP violation. We, however,
may just utilize the existing potential in Ref.\cite{TDLee}, 
replacing the complex scalar fields in the Higgs doublets, by the 
corresponding qauternionic complex scalars defined over 1 and ${\bf i}$.


\end{document}